# Signature Based Detection of User Events for Post-Mortem Forensic Analysis

Joshua Isaac James[1,2], Pavel Gladyshev, and Yuandong Zhu

Centre for Cybercrime Investigation
University College Dublin
Belfield, Dublin 4, Ireland
{Joshua.James, Pavel.Gladyshev, Yuandong.Zhu}@UCD.ie

**Abstract.** This paper introduces a novel approach to user event reconstruction by showing the practicality of generating and implementing signature-based analysis methods to reconstruct high-level user actions from a collection of low-level traces found during a post-mortem forensic analysis of a system. Traditional forensic analysis and the inferences an investigator normally makes when given digital evidence, are examined. It is then demonstrated that this natural process of inferring high-level events from low-level traces may be encoded using signature-matching techniques. Simple signatures using the defined method are created and applied for three popular Windows-based programs as a proof of concept.

## 1 Introduction

The method of using signatures to detect certain types of actions or events is commonplace in many information security related products such as Antivirus and Intrusion Detection Systems (IDS). In these systems, signature based methods have proven to be effective when a known pattern can be tested for. These patterns can range from malicious code embedded in a file, to detecting port scans within a network. "Signatures offer unparalleled percision in detection and forensics… This gives you a clear understanding of exactly what attacks took place… [also] since signatures look for very specific events, they generate a much lower false positive rate…" [3]. The downside, however, is that "traditional signature-based antivirus and antispyware fail to detect zero-day exploits or targeted, custom-tailored attacks" [6], which is a huge disadvantage against todays highly dynamic malware.

---

[1] Research funded by the Science Foundation Ireland (SFI) under Research Frontiers Programme 2007 grant CMSF575.
[2] This Research was conducted using equipment funded by the Higher Education Authority of Ireland under the Research Equipment Renewal Grant Scheme.

Traditional signature-based methods endeavor to detect malicious activities that are currently attempting to access, infect or execute on a system. This paper, however, applies this detection concept in a novel way. The focus is on signatures of user behaviors, where the signature of a *user action* is defined. These signatures are used *after* the incident to detect user actions that have taken place, unlike traditional methods that are used as prevention systems. These signatures are also applied at a system-wide level, looking at the state of the system as a whole, rather than focusing on a single file that may be associated with an activity.

## 2 Contribution

This paper introduces a novel approach to user event reconstruction by showing the practicality of generating and implementing signature-based analysis methods to reconstruct high-level user actions from a collection of low-level traces found during a post-mortem forensic analysis of a system. Specific signatures for common user Windows-based actions are applied as a proof-of-concept for signature-based forensic analysis.

## 3 Organization

This paper begins with an overview of traditional analysis of evidence found within a suspect system. Traditional traces used in an investigation and the inference process that investigators use to derive information from their observations will briefly be discussed. Section five then introduces signature-based detection of user events in an attempt to overcome pitfalls within traditional methods. After which, a short background of Windows timestamps is given where the special case of link file timestamps is also examined. Section seven then demonstrates the method used to derive evidential traces that comprise the basis for the signatures of user actions. Update categories of timestamps are defined which are later used in the creation of these signatures. The signature creation and application process is then shown using Internet Explorer 8 as an example. A demonstration of the application of created signatures is then given using generated signatures for two additional commonly used Windows-based programs. Finally, the results and future of such a technique are considered.

## 4 Traditional Analysis in an Investigation

In traditional digital investigations timestamp information is often used in the analysis phase. Timestamps associated with logs, files, and even registry entries give investigators clues about when certain events took place. However, "use of timestamps as evidence can be questionable due to the reference to a clock with unknown adjustment" [8]. Several methods [5][10][8] to verify the consistency and validity of this valuable timestamp information have been suggested.  Other information, such as inferred event times relative to known events [1][10], can also be derived from traces on a system.

As described in [9][2] the Windows Registry contains much information about user activities. Some, such as Most Recently Used (MRU) lists, contain information that can be directly extracted, while other information, such as the meaning of the order of MRU items through time, must be inferred. Investigators observe these traces of evidence on a system and naturally make inferences as to their meanings based on their knowledge of the system and past experiences. The issue with human inference is that it is manual, prone to cognitive bias/error, and is limited to the amount of knowledge the investigator possesses about the system. "…Our innate inferential abilities are marked by implicit biases that often lead to illogical inference" [4]. These different types of extracted and inferred events have been previously discussed in [10]. The method proposed in this paper attempts to incorporate both extracted and inferred information to automate more of the observed and inferred user event analysis an investigator would normally do during an investigation, with less error and inferential bias.

## 5 Automatic Detection of User Events

When analyzing evidence, investigators normally gather information in two ways: by direct observation and by the inference of one fact from the observation of others. Human inference, however, is prone to assumption and error. To accurately infer information from given facts an investigator must understand the underlying relation between the observed facts and the inferred conclusion. For example Zhu [11] states, "To infer events from the Registry it requires an investigator to understand the relationship between Registry information and occurred activities". This means that when a user does an action that affects data stored in the Windows Registry, the investigator can only begin to infer what the user action was once the investigator understands not only *how* but also *why* that particular piece of data has been modified in the registry.

This paper proposes the theory that both the direct observation and inference phases of an investigation of user actions can be automated using signature-based detection methods. By determining the user activity traces that normally appear in a system after a user event, it is possible to automatically 'infer' the occurrence of the event based on the observable traces. In this paper the focus will be limited to timestamps associated with the files and registry entries. Other traces, such as file fragments in slack space, could also denote user actions, but the scope of this paper is intentionally limited to provide a simple proof of concept. A signature will be defined as a collection of timestamps modified by the occurrence of the event. The hypothesis is that when an event occurs the associated timestamps are updated within a short period of time. As a result, the occurrence of the event may be inferred by observing that the corresponding ensemble of timestamps have been updated within short time of each other. The experiment described in the following sections has been conducted to verify this theory.

## 6 Windows Timestamps

The availability of timestamps differs between different versions of Microsoft Windows. The 'Last Access Time' has been disabled by default for performance reasons in Visa, 2008 and Windows 7. However, "disabling last access update does not mean that the Accessed Date on files does not get updated *at all*; it means that it does not get updated on directory listing or file opening, but last accessed can sometimes be updated when a file is modified and is updated when a file is moved between volumes" [5]. Pre-Vista versions of Windows using the NTFS file system, including Windows 2003, do have last accessed timestamps enabled by default. In all versions with NTFS support the registry setting '*NtfsDisableLastAccessUpdate*'[3] controls if this timestamp is updated. Both modified and created timestamps cannot be disabled with either a simple registry entry or the 'fsutil' utility. Likewise, the Windows Registry provides "[Key Cells] that contain the timestamp of the most recent update to the [Registry] key" [7]. These keys' timestamps can also not be disabled.

### 6.1 Windows Link File Timestamps

Windows Link (.lnk) files are a special case when considering timestamps. According to [5], "when a target file is opened and a link file is created, the created date of the link file remains the date that target file was first accessed during the lifetime of that link file." Parsonage also states, "The Modified Date of the link file represents the time when the related target file was last opened." This information gives valuable insight into user actions since several links to the same target may exist; they may give more information about the target itself; and they are primarily updated on usage rather than execution of the program.

## 7 Deriving User Action Signatures for Internet Explorer

In this section we will discuss the process of deriving signatures for a given user action, in this case "opening Internet Explorer", and show how these signatures can be practically applied. User actions and the concept of causality have previously been discussed in [11]. The same definition of user actions applies, where a user action is an interaction between the user and the system, but in this paper the definition will be applied to the entire system and not only the Windows Registry. The experiment in this section will be conducted on a Windows XP SP3 computer with Internet Explorer 8.0.6001.18702, all with default settings. The first step to defining a signature that can be used to detect a certain user action is to determine the traces that are uniquely updated because of that particular user action.

---

[3] Full path of the last access registry key:
HKEY_LOCAL_MACHINE\SYSTEM\CurrentControlSet\Control\FileSystem\NtfsDisableLastAccessUpdate

### 7.1 Determining Traces for the Signature

Microsoft Process Monitor[4] (procmon) was used to record all system calls executed during the user action "opening Internet Explorer 8 (IE8)". The initial tests recorded all system activity. In order to remove noise (unrelated system calls) generated by other running processes, IE8 was started while recording. This process was done 400 times per test, after which the entries that were not present during every run were removed. Three of these tests were conducted. The filtering process reduced the number of traces from around 11,000 to approximately 4,000 (Fig. 1), however consistent noise was still found to be present.

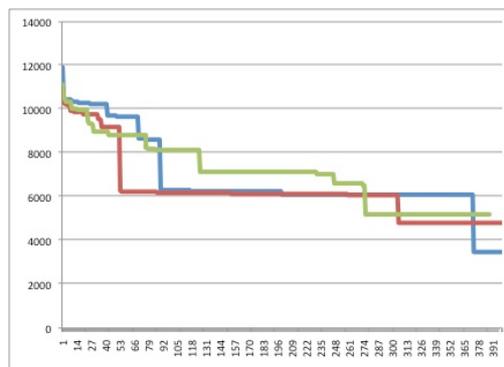

**Fig. 1.** Removal of noise caused by background processes

Using this data it was determined that by filtering the Process Monitor output with the selected program's process name (iexplore.exe), as well as the "explorer.exe" process, similar results of around 4,000 traces could be achieved without needing to repeat the lengthy process multiple times.

To determine the traces associated with starting IE8, Process Monitor was started and configured to filter out events not associated with the "iexplore.exe" and "explorer.exe" processes. The procmon capture was then cleared and IE8 was opened. The procmon capture was stopped and exported as a comma separated value (.csv) file. For ease of processing the file was transferred to a system with various Unix command line tools installed. The following command was used to process the procmon output file "capture.csv":

```
cat capture.csv | awk 'FS=","  { print $5 }' | sort |
uniq > traces.sig
```

This command reads the capture file, removes all the information but the file and registry entry names, sorts these names, removes duplicate entries, and outputs the unique file and registry name list to a file called "traces.sig". The resulting file contains a list of the names and paths of files and registry keys that are accessed during the opening of Internet Explorer 8. It contained 3,915 file and registry traces. Most of which, however, were registry entries that were not Registry keys and therefore have no associated timestamp information. From this list a total of 156 files and 611 registry keys with associated timestamp information were found. A Perl program 'sigtest.pl' was written to retrieve the associated file and registry timestamp information from the items in this list.

---

[4] http://technet.microsoft.com/en-us/sysinternals/bb896645.aspx

To observe the pattern in which the identified trace's timestamps are updated when the user action takes place, Internet Explorer 8 was opened as before. The difference from the first phase is that procmon was not used, and after each opening of IE8, 'sigtest.pl' was executed to output the timestamps of the file and registry entries in the previously created 'traces.sig' file. This process was carried out 10 times over three days. From the produced data, timestamps of the traces can be categorized to determine important event traces as well as update patterns of the final signature.

### 7.2 Analysis of Timestamp Updates

From the data produced in section 7.1, 122 file and registry traces were identified as relevant to the tested user action. This section will list the observed update patterns that will be used to define specific timestamp update categories. Note that, in Windows XP, each file has three associated timestamps, and therefore may count multiple times.

- *Always Updated File and Registry Key Timestamps* (*AU*): It was observed that 21 file and registry timestamps were updated each time IE8 was opened. Of these, 9 files had updated 'accessed' times, 10 files had updated 'modified' times, and 9 registry keys had updated 'modified' times. These timestamps can be further subdivided into five update categories based on the uniqueness of their observed update patterns. These subcategories are explained as follows and are summarized in (Table 1):

Table 1. IE8 *Always Updated* Sub-Category Update Patterns

|     | Modified Time | Accessed Time | Created Time |
|-----|---------------|---------------|--------------|
| *AU1* | Updated | Updated | Unchanged |
| *AU2* | Updated | Updated | Inconsistent |
| *AU3* | Unchanged | Updated | Unchanged |
| *AU4* | N/A | Updated | N/A |
| *AU5* | Updated | Unchanged | Unchanged |

  o  *AU1*: Three files from this group were found to update their accessed and modified timestamps every time Internet Explorer was started, but also with the execution of unrelated actions. Of these files it can be said that their updated timestamps must be greater-than or equal to the time of the most recent execution of IE8. It was also observed that the created timestamps of these files is less-than or equal to the installation of the system itself.

- *AU2*: Three files were observed to have their accessed and modified timestamps updated with each execution of IE8. Of these files, one was the prefetch[5] file for Internet Explorer. Its created timestamp was found to correlate with the first time Internet Explorer was run on the system. Only the accessed and modified times were updated with each user action. The other two files were Internet Explorer 'cookie' files correlating to 'administrator@live[1].txt' and 'administrator@msn[1].txt' where 'administrator' is the name of the local user account. The accessed and modified timestamps of these files were updated with each user action, and the created timestamps were found to update often with the user action, but not always and with no discernable pattern. Of these files it can be said that any timestamp happening before the most recent user event denotes the time of a previous user event.

- *AU3*: Four files were found to have only their modified, and not their accessed, times updated with each opening of Internet Explorer 8.

- *AU4*: Nine Registry keys were identified that always had their associated timestamp information updated.

- *AU5*: Two files were found to have only it's accessed timestamps updated: IExplore.exe and shell32.dll.

- *Timestamps Updated on the First Run Only* (*FRO*): It was observed that 1 registry timestamp was updated only during the first opening of IE8 per session.

- *Usage-Based File Timestamp Updates* (*UB*): It was observed that 4 Windows shortcut (.lnk) files' accessed timestamps were updated often, but not always when they were used to start IE8. If they were not used to start IE8 they were never updated by the action.

- *Irregular Update of Timestamps* (*IU*): It was observed that 93 files had irregular timestamp update patterns, and each in this category had only its accessed timestamp updated.

  - *IU1*: Although the majority of the traces categorized as IU are seemingly irregular, it was observed that cookies within the user's "\Cookies" folder were updated on the first run of the session, and then irregularly updated during the starting of IE8 in the same user session, making cookie files a combination of *FRO* and *IU*.

---

[5] More information about prefetch files can be found at: http://www.microsoft.com/whdc/archive/XP_kernel.mspx#ECLAC

8      **Joshua Isaac James** , , Pavel Gladyshev, and Yuandong Zhu

**7.3 Categories of Timestamps**

From the observed update patterns, four primary categories of timestamp updates can be defined.

There are two important observations that apply to each category. First, trace updates are *caused* by a user action, such as double-clicking an icon. This process is not instantaneous and therefore any observable traces were created or updated some time *after* the actual user action. Second, it was observed that each trace was updated within one minute of the user action. This means that the update process must also be defined as a *time-span* and is not instantaneous.

Category 1: *Always Updated Timestamps* – 6 files and registry entries with timestamp information were consistently updated each time, and only when, Internet Explorer 8 was opened. Traces that are always updated by opening IE8, as well as by other user actions, have been removed. The remaining traces in this category will provide the core of the signature, as they are the most reliably updated.

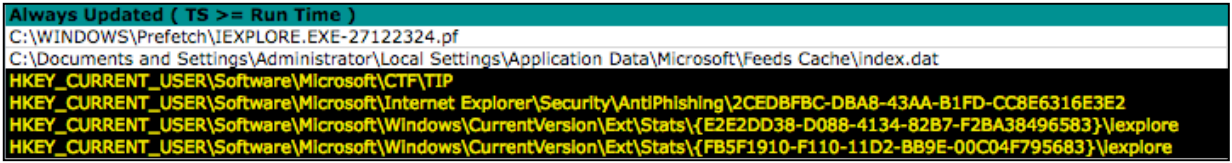

**Fig. 2.** IE8 file and registry traces updated each time IE8 is opened

Category 2: *Timestamps Updated on the First Run Only* – One registry entry and all cookie files were found to have their timestamp information consistently updated on the first run of Internet Explorer 8 *per user session* (Fig. 3).

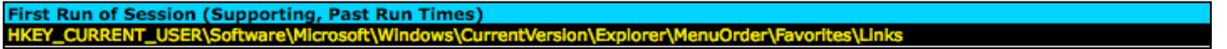

**Fig. 3.** IE8 registry trace updated during the first run of the session

Category 3: *Irregular Update of Timestamps* – 93 files were found to have their timestamps updated in an irregular fashion (Appendix A).

Category 4: *Usage-Based Timestamp Update* – 4 Windows shortcut files were identified that were inconsistently updated when the particular link file itself was used (Fig. 4), and never during the starting of IE8 when the file itself was not used.

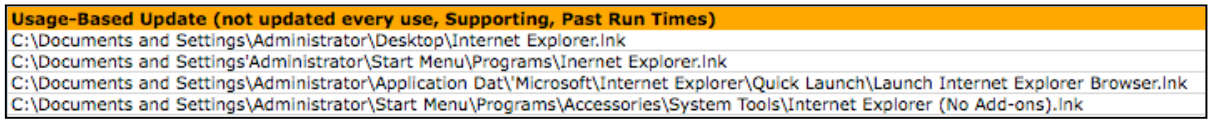

**Fig. 4.** IE8 file traces updated on file usage (link files)

Also from the original traces list there were a number of file and registry entries that were never updated during the opening of IE8. These entries have been discarded.

By using the different categories of timestamps previously explained, signatures that match the known timestamp update patterns can now be derived.

### 7.4 Signature Generalization

Once a list of traces associated with a certain user action is generated and classified, the traces must then be generalized to be portable. To do this, any user or system-specific paths would have to be generalized. Taking the Category 1 Windows prefetch file as an example:

C:\Windows\Prefetch\IEXPLORE.EXE-27122324.pf

The system-unique identifiers would need to be replaced with variables, as so:

**%SystemRoot%**\Prefetch\IEXPLORE.EXE-**%s**.pf

Where the variable %SystemRoot% is the location of the Windows system folder including the drive and path, and %s is a string of numbers and letters.

The generalization should include the possibility that programs may be installed in non-default locations. This means that other information sources, such as reading the installation path from the Windows Registry[6], would be required. This generalization will allow signatures generated on one system to be used in the analysis of other computers.

### 7.5 Creation of the Signature for Opening Internet Explorer

A signature in the context of this paper is defined as a collection of timestamps modified by the occurrence of an event. By using the previously defined categories of timestamps as well as the observations of update patterns within these categories, a signature of a particular user action can be created by defining the pattern in which associated trace timestamps are updated during the occurrence of the user action.

Category 1 traces will provide the Core of the created signatures. The reason for this is that the modified timestamps in this group must always be updated given a user opening IE8. Due to the inconsistent nature of the other category types, they will be defined as 'supporting evidence'. Supporting evidence can enhance the probability, or believability, of a detected user action, but the issue of probability is beyond the scope of this paper.

Based on observations relating to the always-updated file and registry key timestamps (timestamp Category 1), a signature can be defined. Using traces that are always updated *only* by opening IE8, the execution of the user action can be inferred from these traces if all the traces display consistent update information. In this case consistency means that each trace has a timestamp that has been updated within 1 minute of each other. If the traces are not consistent with each other, then nothing can be inferred since some unknown, uncommon action must have updated the traces in an unexpected way. Consider the Core traces from IE8 (generalized):

- %SystemRoot%\Prefetch\IEXPLORE.EXE-%s.pf
- %HomeDrive%\%HomePath%\Local Settings\Application Data\Microsoft\Feeds Cache\index.dat
- HKEY_USERS\%SID\Software\Microsoft\CTF\TIP
- HKEY_USERS\%SID\Software\Microsoft\Internet Explorer\Security\AntiPhishing\%s
- HKEY_USERS\%SID\Software\Microsoft\Windows\CurrentVersion\Ext\Stats\{%s}\iexplore

---

[6] One Windows Registry key containing installed program path information is:
HKEY_LOCAL_MACHINE\Software\Microsoft\Windows\CurrentVersion\Uninstall

These traces are updated every time, and only, when the user action "open Internet Explorer 8" is executed. The collection of modified timestamps of the detected traces may then be tested for consistency: whether each timestamp has been updated within 1 minute of each other. This process is shown below.

For the first test IE8 was executed on a Windows XP system at 2:30pm on the 12$^{th}$ of April 2010. Various other programs such as Mozilla Firefox, Windows Live Messenger, Outlook Express, and others were used to browse the Internet, chat and check email, respectively. These actions took place over two days without another execution of IE8. During the evenings the computer was shut down, and was restarted the next morning. The following table (Table 2) shows the timestamps of the IE8 traces analyzed on the 14$^{th}$ of April 2010.

Table 2. "Open IE8" Signature Analysis conducted at 4:45pm 14/4/2010

| Trace Name | Timestamp |
| --- | --- |
| C:\WINDOWS\Prefetch\IEXPLORE.EXE-27122324.pf | 4/12/2010 14:30:37 |
| C:\Documents and Settings\Administrator\Local Settings\Application Data\Microsoft\Feeds Cache\index.dat | 4/12/2010 14:30:26 |
| HKEY_USERS\S-1-5-21-1417001333-573735546-682003330-500\Software\Microsoft\CTF\TIP | 4/12/2010 2:30 PM |
| HKEY_USERS\S-1-5-21-1417001333-573735546-682003330-500\Software\Microsoft\Internet Explorer\Security\AntiPhishing\2CEDBFBC-DBA8-43AA-B1FD-CC8E6316E3E2 | 4/12/2010 2:30 PM |
| HKEY_USERS\S-1-5-21-1417001333-573735546-682003330-500\Software\Microsoft\Windows\CurrentVersion\Ext\Stats\{E2E2DD38-D088-4134-82B7-F2BA38496583}\iexplore | 4/12/2010 2:30 PM |
| HKEY_USERS\S-1-5-21-1417001333-573735546-682003330-500\Software\Microsoft\Windows\CurrentVersion\Ext\Stats\{FB5F1910-F110-11D2-BB9E-00C04F795683}\iexplore | 4/12/2010 2:30 PM |

Each trace has a timestamp that was updated within 1 minute of 2:30pm, and all correlate to the time Internet Explorer was last opened. Next IE8 was opened again on the 14[th] of April 2010 at 5:00pm. The trace timestamps were analyzed at 5:19pm, the result of which is shown in table 3.

Table 3. "Open IE8" Signature Analysis conducted at 5:19pm 14/4/2010

| Trace Name | Timestamp |
|---|---|
| C:\WINDOWS\Prefetch\IEXPLORE.EXE-27122324.pf | 4/14/2010 17:00:24 |
| C:\Documents and Settings\Administrator\Local Settings\Application Data\Microsoft\Feeds Cache\index.dat | 4/14/2010 17:00:19 |
| HKEY_USERS\S-1-5-21-1417001333-573735546-682003330-500\Software\Microsoft\CTF\TIP | 4/14/2010 5:00 PM |
| HKEY_USERS\S-1-5-21-1417001333-573735546-682003330-500\Software\Microsoft\Internet Explorer\Security\AntiPhishing\2CEDBFBC-DBA8-43AA-B1FD-CC8E6316E3E2 | 4/14/2010 5:00 PM |
| HKEY_USERS\S-1-5-21-1417001333-573735546-682003330-500\Software\Microsoft\Windows\CurrentVersion\Ext\Stats\{E2E2DD38-D088-4134-82B7-F2BA38496583}\iexplore | 4/14/2010 5:00 PM |
| HKEY_USERS\S-1-5-21-1417001333-573735546-682003330-500\Software\Microsoft\Windows\CurrentVersion\Ext\Stats\{FB5F1910-F110-11D2-BB9E-00C04F795683}\iexplore | 4/14/2010 5:00 PM |

Each trace has a timestamp that was updated within 1 minute of 5:00pm, and all correlate to the time Internet Explorer was last opened.
    Since these timestamps must be updated when the user action takes place, and the updates are caused only by the user action, then if all these timestamps are consistent then it can be inferred that the user action that caused the updates took place shortly before the detected timestamps.

### 7.6 Further Application of Signatures

To determine whether this approach is applicable to programs other than IE8, signatures of user actions for Firefox 3.6 and MSN Messenger 2009 were created using the process described earlier.

### 7.6.1 Detecting the Opening of Firefox 3.6

For Firefox 3.6 (FF3.6) the user action of "opening Firefox 3.6" was tested. 1,507 original traces were updated when FF3.6 was opened. Of these, only 1 file was determined to have always updated (Category 1) timestamps. This file was the standard Windows pre-fetch file:

C:\WINDOWS\Prefetch\FIREFOX.EXE-28641590.pf

Other traces were identified as being updated when FF3.6 was opened, but were also updated when FF3.6 was used to browse the Internet or when the program was closed. Because only one trace is available other sources of user event information will need to be identified to ensure this trace is consistent with the system, thereby increasing the reliability of the observed and inferred information.

**7.6.2 Detecting the Opening of MSN Messenger 2009**

For MSN Messenger 2009 (MSN2009) the user action of "opening MSN Messenger 2009" was tested. 4,263 original traces were updated when MSN2009 was opened. Of these, 3 unique traces were determined to have always updated (Category 1) timestamps.

- %HomeDrive%\%HomePath%\Tracing\WindowsLiveMessenger-uccapi-%i.uccapilog
- %SystemRoot%\Prefetch\MSNMSGR.EXE-%s.pf
- HKEY_USERS\%SID\Software\Microsoft\Tracing\WPPMedia

To test this signature MSN2009 was started at 7:28pm on the 14$^{th}$ of April 2010. At 7:29pm MSN2009 was closed. IE8, FF3.6 and Outlook Express were then used to surf the Internet and check email. The signature analysis was then conducted at 7:49pm on the same day. The results of the analysis are shown in table 4.

Table 4. "Open MSN2009" Signature Analysis conducted at 7:49pm 14/4/2010

| Trace Name | Timestamp |
|---|---|
| C:\Documents and Settings\Administrator\Tracing\WindowsLiveMessenger-uccapi-0.uccapilog | 4/14/2010 19:28:25 |
| C:\WINDOWS\Prefetch\MSNMSGR.EXE-030AB647.pf | 4/14/2010 19:28:25 |
| HKEY_USERS\S-1-5-21-1417001333-573735546-682003330-500\Software\Microsoft\Tracing\WPPMedia | 4/14/2010 7:28 PM |

Each trace has a timestamp that was updated within 1 minute of 7:28pm, and all correlate to the time MSN2009 was last opened. Next MSN2009 was opened again on the 14$^{th}$ of April 2010 at 7:56pm. The trace timestamps were analyzed at 7:58pm, the result of which is shown in table 5.

Table 5. "Open MSN2009" Signature Analysis conducted at 7:58pm 14/4/2010

| Trace Name | Timestamp |
|---|---|
| C:\Documents and Settings\Administrator\Tracing\WindowsLiveMessenger-uccapi-0.uccapilog | 4/14/2010 19:56:46 |
| C:\WINDOWS\Prefetch\MSNMSGR.EXE-030AB647.pf | 4/14/2010 19:56:46 |
| HKEY_USERS\S-1-5-21-1417001333-573735546-682003330-500\Software\Microsoft\Tracing\WPPMedia | 4/14/2010 7:56 PM |

Each trace has a timestamp that was updated within 1 minute of 7:56pm, and all correlate to the time MSN2009 was last opened. It can be inferred that the user action "open MSN2009" must have taken place shortly before the detected timestamps.

## 8. Conclusions

Traditional analysis in a digital investigation is currently a highly manual process. With the growing amount of data an investigator must analyze, automated analysis techniques are necessary. This paper demonstrated how signature-based detection methods could be used to detect defined user actions by inferring information from the patterns in which traces are updated for the given user action. A simple signature for a particular user action has been created and applied to automatically detect the last occurrence of a user action during a post-mortem investigation. Even though detection of simple user actions for three programs has been shown, this technique does not fully utilize all the observable information, requires much more extensive testing across many systems, and has yet to demonstrate its practicality for the detection of more complex user actions. For these reasons there is still much work to be done.

## 9. Future Work

Based on the results obtained in this paper it appears that signature-based detection of user actions is possible, however much work needs to be done. Requirements include making the signatures portable, for example detecting traces within the signatures if the traces have been installed in different (non-default) locations, or even the use of this method on other operating systems. Others include improving the usage of information gained from observable traces, i.e. what other information can be inferred. This area includes the introduction of probability by attempting to capitalize on the 'supporting evidence' defined earlier. Also the detection of 'exact matches' within the Core signature may prove to provide supporting information that may help ensure the consistency and integrity of the observed, and thus inferred, information. One final consideration is the ability to detect not only the most recent time a user action has happened, but also previous executions of the user action based on the observable information.

## References

1. Gladyshev, Pavel, Ahmed Patel. Formalising Event Time Bounding in Digital Investigations. *International Journal of Digital Evidence,* vol. 4. 2005.
2. Kahvedzic, Damir, Tahar Kechadi. Extraction of user activity through comparison of windows restore points. *6th Australian Digital Forensics Conference*. 2008.
3. McAfee. Complete Security: The Case for Combined Behavioral and Signature-Based Protection. Whitepaper. Santa Carla: McAfee Inc., 2005.
4. Ogawa, Akitoshi, Yumiko Yamazaki, Kenichi Ueno, Kang Cheng, and Atsushi Iriki. Neural Correlates of Species-typical Illogical Cognitive Bias in Human Inference. *Journal of Cognitive Neuroscience*. Massachusetts Institute of Technology. 2009. doi:10.1162/jocn.2009.21330.
5. Personage, Harry. The Meaning of (L)inkfiles (I)n (F)orensic (E)xaminations. November 2009. Computer Forensics Miscellany. 2 Febuary 2010. <http://computerforensics.parsonage.co.uk/downloads/TheMeaningofLIFE.pdf>.
6. Roiter, Neil. When signature based antivirus isn't enough. 3 May 2007. 2 Febuary 2010 <http://searchsecurity.techtarget.com/news/article/0,289142,sid14_gci1253602,00.html>.
7. Russinovich, Mark. Inside the Registry. 3 Feburary, 2010. <http://technet.microsoft.com/en-us/library/cc750583.aspx>.
8. Willassen, Svein Y. Timestamp evidence correlation by model based clock hypothesis testing. *Proceedings of the 1st international Conference on Forensic Applications and Techniques in Telecommunications, information, and Multimedia and Workshop*, ICST, Brussels, Belgium, 1-6. 2008.
9. Zhu, Yuandong, Joshua James, Pavel Gladyshev. A comparative methodology for the reconstruction of digital events using Windows Restore Points. Digital Investigation. 2009a; doi:10.1016/j.diin.2009.02.004.
10. Zhu, Yuandong, Joshua James, Pavel Gladyshev. Consistency Study of the Windows Registry. *Sixth Annual IFIP WG 11.9 International Conference on Digital Forensics*. 2010.
11. Zhu, Yuandong, Pavel Gladyshev, Joshua James. Using ShellBag Information to Reconstruct User Activities. *Digital Investigation*, vol. 6. 69-77. 2009c; doi:10.1016/j.diin.2009.06.009

## Appendix

**Appendix A**: List of Category 3 Traces for Internet Explorer– Irregularly Updated.

C:\Documents and Settings\Administrator\Cookies\*,
C:\WINDOWS\system32\ieapfltr.dat,
C:\Documents and Settings\Administrator\Application Data\Microsoft\IdentityCRL\production\ppcrlconfig.dll,
C:\Documents and Settings\All Users\Application Data\Microsoft\IdentityCRL\production\ppcrlconfig.dll,
C:\Documents and Settings\Administrator\Application Data\Microsoft\CryptnetUrlCache\Content\7B2238AACCEDC3F1FFE8E7EB5F575EC9,
C:\Documents and Settings\Administrator\Application Data\Microsoft\CryptnetUrlCache\MetaData\7B2238AACCEDC3F1FFE8E7EB5F575EC9,
C:\WINDOWS\system32\xmllite.dll,
C:\Documents and Settings\Administrator\Local Settings\Application Data\Microsoft\Internet Explorer\frameiconcache.dat,

C:\Documents and Settings\Administrator\Favorites\Links\desktop.ini,
C:\Documents and Settings\Administrator\Favorites\Desktop.ini,
C:\WINDOWS\system32\winhttp.dll,
C:\Program Files\Common Files\Microsoft Shared\Windows Live\WindowsLiveLogin.dll,
C:\Program Files\Common Files\Microsoft Shared\Windows Live\msidcrl40.dll
C:\WINDOWS\system32\ieui.dll
C:\WINDOWS\system32\msls31.dll
C:\WINDOWS\system32\ieapfltr.dll
C:\Program Files\Internet Explorer\xpshims.dll
C:\WINDOWS\system32\mshtml.dll
C:\WINDOWS\system32\msfeeds.dll
C:\WINDOWS\system32\activeds.dll
C:\WINDOWS\system32\adsldpc.dll
C:\WINDOWS\system32\credui.dll
C:\WINDOWS\system32\cryptnet.dll
C:\WINDOWS\system32\cscdll.dll
C:\WINDOWS\system32\cscui.dll
C:\WINDOWS\system32\dhcpcsvc.dll
C:\WINDOWS\system32\dot3api.dll
C:\WINDOWS\system32\dot3dlg.dll
C:\WINDOWS\system32\eapolqec.dll
C:\WINDOWS\system32\eappcfg.dll
C:\WINDOWS\system32\eappprxy.dll
C:\WINDOWS\system32\esent.dll
C:\WINDOWS\system32\mprapi.dll
C:\WINDOWS\system32\msxml3r.dll
C:\WINDOWS\system32\netman.dll
C:\WINDOWS\system32\netshell.dll
C:\WINDOWS\system32\onex.dll
C:\WINDOWS\system32\psapi.dll
C:\WINDOWS\system32\qutil.dll
C:\WINDOWS\system32\rasadhlp.dll
C:\WINDOWS\system32\rsaenh.dll
C:\WINDOWS\system32\winlogon.exe
C:\WINDOWS\system32\winrnr.dll
C:\WINDOWS\system32\wintrust.dll
C:\WINDOWS\system32\wmi.dll
C:\WINDOWS\system32\wtsapi32.dll
C:\WINDOWS\system32\wzcsapi.dll
C:\WINDOWS\system32\wzcsvc.dll
C:\Program Files\Messenger\msmsgs.exe
C:\WINDOWS\system32\mswsock.dll
C:\WINDOWS\system32\msxml3.dll
C:\WINDOWS\system32\atl.dll
C:\Program Files\Internet Explorer\sqmapi.dll
C:\WINDOWS\system32\schannel.dll
C:\WINDOWS\AppPatch\aclayers.dll
C:\WINDOWS\system32\urlmon.dll
C:\Program Files\Internet Explorer\ieproxy.dll
C:\WINDOWS\system32\iertutil.dll
C:\WINDOWS\system32\ieframe.dll
C:\WINDOWS\system32\actxprxy.dll
C:\WINDOWS\system32\apphelp.dll
C:\WINDOWS\system32\crypt32.dll
C:\WINDOWS\system32\cryptdll.dll
C:\WINDOWS\system32\digest.dll
C:\WINDOWS\system32\iphlpapi.dll
C:\WINDOWS\system32\ir32_32.dll
C:\WINDOWS\system32\ir41_32.ax
C:\WINDOWS\system32\ir41_qc.dll
C:\WINDOWS\system32\ir41_qcx.dll
C:\WINDOWS\system32\ir50_32.dll
C:\WINDOWS\system32\ir50_qc.dll
C:\WINDOWS\system32\ir50_qcx.dll
C:\WINDOWS\system32\mlang.dll
C:\WINDOWS\system32\msapsspc.dll
C:\WINDOWS\system32\msisip.dll
C:\WINDOWS\system32\msnsspc.dll
C:\WINDOWS\system32\msvcrt40.dll
C:\WINDOWS\system32\rasapi32.dll
C:\WINDOWS\system32\rasman.dll
C:\WINDOWS\system32\rtutils.dll
C:\WINDOWS\system32\sensapi.dll
C:\WINDOWS\system32\setupapi.dll
C:\WINDOWS\system32\sxs.dll
C:\WINDOWS\system32\tapi32.dll
C:\WINDOWS\system32\winspool.drv
C:\WINDOWS\system32\ws2_32.dll
C:\WINDOWS\system32\ws2help.dll
C:\WINDOWS\system32\xpsp2res.dll
C:\WINDOWS\system32\msv1_0.dll
C:\WINDOWS\system32\msasn1.dll
C:\WINDOWS\system32\wshext.dll
C:\WINDOWS\system32\dnsapi.dll
C:\Documents and Settings\Administrator\Cookies\administrator@live[1].txt
C:\Documents and Settings\Administrator\Cookies\administrator@msn[1].txt